\documentclass[aps,superscriptaddress,twocolumn,prl]{revtex4-2}

\usepackage{bm,graphicx,textcomp,amssymb,amsmath,dcolumn,color}
\usepackage[normalem]{ulem}

\newcommand{\ket}[1]{\left|#1\right>}
\newcommand{\bra}[1]{\left<#1\right|}
\newcommand{\braket}[2]{\left<#1|#2\right>}

\newcommand{\f}[1]{\mbox{\boldmath$#1$}}

\newcommand{\bea}{\begin{eqnarray}}
\newcommand{\ea}{\end{eqnarray}}
\newcommand{\eea}{\end{eqnarray}}
\newcommand{\ord}{\,{\cal O}}

\begin{document}

\title{Classical versus non-classical photon states
for detecting 
vacuum non-linearity}

\author{N.~Ahmadiniaz} 
\affiliation{Helmholtz-Zentrum Dresden-Rossendorf,
Bautzner Landstra\ss e 400, 01328 Dresden, Germany}

\author{C. Kohlf\"urst}
\affiliation{Helmholtz-Zentrum Dresden-Rossendorf,
Bautzner Landstra\ss e 400, 01328 Dresden, Germany}

\author{R.~Shaisultanov}
\affiliation{The Extreme Light Infrastructure ERIC, 
	ELI Beamlines Facility, Za Radnicí 835, 252 41 Doln\`{i} B\v{r}e\v{z}any, Czech Republic}

\author{R.~Sch\"utzhold}
\affiliation{Helmholtz-Zentrum Dresden-Rossendorf,
Bautzner Landstra\ss e 400, 01328 Dresden, Germany}
\affiliation{Institut f\"ur Theoretische Physik,
Technische Universit\"at Dresden, 01062 Dresden, Germany}

\begin{abstract}
Quantum electrodynamics (QED) predicts that the vacuum should behave as
a non-linear medium.
For many schemes aimed at verifying this fundamental prediction,
the signal consists of one or more photons in a certain mode
(determined by polarization $\sigma$ and wave-number $\f{k}$)
which would be empty in the absence of a QED vacuum non-linearity.
Here we consider modifying these schemes by sending in a classical
or non-classical (e.g., squeezed) photon state instead of the initial
vacuum state in that mode.
We find that the detectability can be improved significantly.
\end{abstract}

\date{\today}

\maketitle

\paragraph{Introduction}

Combining the electron mass $m$ with the elementary charge $q$ 
yields fundamental scales of quantum electrodynamics (QED) 
such as the Compton length 
$\lambdabar=\hbar/(mc)\approx386~\rm fm$ 
or the Schwinger critical field strength 
$E_{\rm crit}=m^2c^3/(q\hbar)\approx1.3\times10^{18}~\rm V/m$ \cite{HeisenbergEuler1936,Schwinger1951}.
In order to test QED in various regimes, one route is to study short 
distances, i.e., high energies and momenta -- but even on long distances
$\gg\lambdabar$, another route is to investigate high field strengths. 
In this regime, QED predicts that the vacuum behaves as a non-linear 
medium featuring phenomena such as 
four-wave mixing or birefringence, see, e.g.,
\cite{HeisenbergEuler1936,Schwinger1951, Karplus:1950zza,
Karplus:1950zz, Baier, Bialynicka-Birula:1970nlh, DiPiazza2012,
Fedotov2023}.

In spite of strong efforts, however, the predicted vacuum non-linearity
on long distances 
has not been conclusively verified yet~\cite{Ejlli2020,Fedotov2023}.
The considered scenarios include birefringence 
(photon polarization rotation) in strong static magnetic
fields~\cite{Ejlli2020},
four-wave mixing with strong optical
lasers~\cite{Lundin2006,Tommasini2010,PhysRevD.109.096009,
Rinderknecht:2025zgu} or x-ray birefringence in strong optical lasers~\cite{Heinzl2006,Karbstein2021,Karbstein2022PRL,
AhmadiniazDetection2023,BIREF-HIBEF2025,Shen:2018lbq,Bragin:2017yau}.
In most of these set-ups, the idea is that the vacuum non-linearity 
creates one or more photons in a certain mode 
(determined by polarization $\sigma$ and wave-number $\f{k}$,
i.e., frequency and propagation direction)
where there would be no photons without a
non-linearity~\cite{BIREF-HIBEF2025,Smid2025}.
In other words, the initial state in that specific mode is the vacuum
state.

For sensing in quantum optics, for example, it is well known that  one
may improve the sensitivity if, instead of the incoming vacuum state, 
other states are prepared as the initial
state~\cite{Caves1981,Giovannetti2004,Giovannetti2011,LIGO2013}.
In the following, we investigate how this idea could be transferred 
to the detection of the quantum vacuum non-linearity and study how 
various classical and non-classical initial states could facilitate 
its detection. 

\paragraph{Formalism}

For fields strengths well below the Schwinger critical field
$E_{\rm crit}$ and energy-momentum scales far below the electron mass $m$,
the quantum vacuum non-linearity can be described by the lowest-order 
Euler-Heisenberg Lagrangian ($\hbar=c=\varepsilon_0=\mu_0=1$)
\bea
\label{Euler-Heisenberg}
{\cal L}=\frac12\left(\f{E}^2-\f{B}^2\right)+
\xi\left[\left(\f{E}^2-\f{B}^2\right)^2+7\left(\f{E}\cdot\f{B}\right)^2
\right],
\ea
with the coupling constant $\xi=q^4/(360\pi^2m^4)$
\cite{HeisenbergEuler1936,Schwinger1951}.
%
%
Splitting the Lagrangian~\eqref{Euler-Heisenberg} into 
${\cal L}_0=(\f{E}^2-\f{B}^2)/2$ and the interaction term 
${\cal L}_{\rm int}$, we employ the interaction picture 
and arrive at the time evolution operator 
\bea
\label{evolution}
\hat U
=
{\cal T}\exp\left\{-i\int d^4x\;\hat{\cal H}_{\rm int}\right\}, 
\ea
where the interaction Hamiltonian $\hat{\cal H}_{\rm int}$ is proportional 
to the small coupling constant $\xi$ and of fourth order in the field 
strengths. 
As the next step, we split up all fields 
$\hat A^\mu=A^\mu_{\rm ext}+\delta\hat A^\mu$ 
into a large classical background field $A^\mu_{\rm ext}$
plus small quantum fluctuations $\delta\hat A^\mu$.
%
To obtain a detectable signal, the smallness of $\xi$ must be 
compensated by large background fields $A^\mu_{\rm ext}$
(albeit still well below the Schwinger limit
$E_{\rm crit}$).
Thus, we keep only terms to linear order in small contributions $\xi$ 
and $\delta\hat A^\mu$, 
i.e., cubic terms in the 
large classical background field $A^\mu_{\rm ext}$. 

In this lowest order, the time ordering $\cal T$ in Eq.~\eqref{evolution} 
corresponds to commutators of $\delta\hat A^\mu$ 
(i.e., bi-linear terms) and can thus be omitted. 
Altogether, we can approximate the time evolution
operator~\eqref{evolution} by
\bea
\label{effective}
\hat U_{\rm eff}
=
\exp\left\{i\int d^4x\;J_\mu^{\rm vac}\delta\hat A^\mu\right\}
=
\exp\left\{{\mathfrak A}\,\hat a^\dagger-\rm h.c.\right\}, \quad
\ea
where $\hat a^\dagger$ is the creation operator for the effective mode
of the electromagnetic field $\delta\hat A^\mu$ 
after inserting the classical background field $A^\mu_{\rm ext}$ 
and performing the space-time integration. 
The effective current $J_\mu^{\rm vac}$ encodes the vacuum non-linearity
and is given by a cubic form of the classical background field
$A^\mu_{\rm ext}$ (vacuum emission picture, see also  \cite{KarbsteinShaisultanov2015,GieKarKoh2018}).

In order to distinguish the signal mode $\hat a^\dagger$ and $\hat a$ 
from the background $A^\mu_{\rm ext}$,  
we focus on those modes $\hat a^\dagger$ and $\hat a$ which propagate 
into a different direction and/or have a different polarization than 
$A^\mu_{\rm ext}$ 
(more explicit experimental schemes will be discussed below).
%

\paragraph{Vacuum state}

In most set-ups for detecting the quantum vacuum non-linearity, 
the signal mode 
starts in the vacuum 
state $\ket{0}$ such that its final state is given by
\bea
\label{vacuum}
\ket{{\rm out}}=\hat U_{\rm eff}\ket{0}
=\exp\left\{{\mathfrak A}\,\hat a^\dagger-\rm h.c.\right\}
\ket{0}
\,,
\ea
i.e., we obtain a coherently displaced vacuum
state~\cite{Glauber1963,AhmadiniazDetection2020}.
For a detection of quantum vacuum non-linearity~\eqref{effective}, 
it is necessary to distinguish this final state~\eqref{vacuum} from the 
initial unperturbed state $\ket{{\rm in}}=\ket{0}$.

In the small-signal limit where 
$|{\mathfrak A}|\ll1$, the final state is given 
by $\ket{{\rm out}}=\ket{0}+{\mathfrak A}\ket{1}+\ord({\mathfrak A}^2)$
and thus has a large overlap with the initial vacuum state. 
As a result, it can typically not be distinguished from the initial state
in a single run, i.e., many repetitions of the experiment are required
\cite{Caves1981,Giovannetti2004,Giovannetti2011}.
A simple and very direct way of distinguishing would be to measure the
final photon number, which then yields a non-zero value with probability 
$\approx|{\mathfrak A}|^2$.

\paragraph{Coherent state}

For a given set-up, we know (at least in principle) the signal mode 
$\hat a^\dagger$ and $\hat a$. 
Thus, instead of the vacuum state, one could also send in another state 
in this mode, such as a classical, i.e., coherent state \cite{Glauber1963}
%
\bea
\ket{{\rm in}}=\ket{\alpha}=\hat D(\alpha)\ket{0}
=\exp\left\{\alpha\hat a^\dagger-\rm h.c.\right\}
\ket{0}
\,.
\ea
Since the norm of the states $\hat a^\dagger\ket{\alpha}$ and
$\hat a\ket{\alpha}$ can be much larger than unity for large
displacements $|\alpha|\gg1$, the interaction~\eqref{effective}
can be enhanced in this way (e.g., regarding the amount of energy
exchanged). 

However, this enhancement in itself does not imply an improved 
detectability, which requires distinguishing the states $\ket{\alpha}$ 
and $\hat U_{\rm eff}\ket{\alpha}$. 
Since $\hat U_{\rm eff}$ and $\hat D(\alpha)$ are both coherent 
displacements, we may use the property \cite{MandelWolf}
\bea
\label{addition}
\hat U_{\rm eff}({\mathfrak A})\hat D(\alpha)
=
\exp\left\{\frac{{\mathfrak A}\alpha^*-{\mathfrak A}^*\alpha}{2}\right\}
\hat D({\mathfrak A}+\alpha)
\,,
\ea
i.e., the result is again a coherent state. 
Since two different coherent states are not orthogonal but have a non-zero 
overlap $|\braket{\alpha_1}{\alpha_2}|^2=\exp\{-|\alpha_1-\alpha_2|^2\}$, 
we see that the detectability of the vacuum non-linearity~\eqref{effective},
i.e., the ability to distinguish the states $\ket{{\rm in}}=\ket{\alpha}$
and $\ket{{\rm out}}=\hat U_{\rm eff}\ket{\alpha}$, 
again requires an amplitude ${\mathfrak A}$ of order unity (or above) 
and is thus basically the same as in the vacuum case. 

However, we may also try to exploit the phase in Eq.~\eqref{addition}. 
%
If $\alpha$ and $\mathfrak A$ have the same phase (e.g., both are real),
the interaction~\eqref{effective} would just change the mean energy or 
photon number of the incident state $\ket{\alpha}$.
If, on the other hand, the phases of $\alpha$ and $\mathfrak A$
are shifted by $\pi/2$ (e.g., one is real and one imaginary), 
then the interaction~\eqref{effective} basically corresponds 
to a phase shift. 
In view of the mean number of photons $n=|\alpha|^2$ in this coherent 
state, the total phase shift scales with ${\mathfrak A}\sqrt{n}$ 
such that the phase shift per photon scales as ${\mathfrak A}/\sqrt{n}$. 
If we now compare this shift to the well-known Poisson limit for phase 
accuracy $\Delta\varphi\sim1/\sqrt{n}$, we again find that an amplitude
${\mathfrak A}$ of order unity or larger is required for
detectability (at least in a single run), see also  \cite{AhmadiniazDetection2020}.

Nevertheless, even though coherent states do not offer any enhancement 
in this respect, they could still be advantageous
for detecting the signal in the presence of stray (i.e., noise)
photons from other sources.
In the presence of many stray photons, detecting a single signal photon
-- as in the case of the initial vacuum state -- reliably may be very
hard, but measuring a phase shift of a coherent state $\ket{\alpha}$
(e.g., by interference with a reference beam) containing many photons
can be far more robust.

Note that this scheme depends on the relative phase between the signal mode 
and the incident coherent state, which may be hard to determine in practice. 
However, even for random phases, one should see a corresponding
signal in the interference patterns after several repetitions of
the experiment.
Nevertheless, coherent states remain limited by the Poisson
(or shot-noise) limit, motivating the use of non-classical probe
states discussed below.

\paragraph{Single-mode squeezed state}

In quantum optics, a well-known state which allows us to enhance the 
sensitivity is a squeezed state \cite{Yuen1976,Caves1981}.
\bea
\label{single-mode}
\ket{\zeta}=
\hat S_\zeta\ket{0}=
\exp\left\{\frac{\zeta^*}{2}\,\hat a^2-\rm h.c.\right\}
\ket{0}.
\ea
In such a squeezed state, the variance or uncertainty in one direction 
in phase space is reduced while it is correspondingly enlarged in the 
other direction.
This reduced variance can then help us to detect small displacements
from the interaction~\eqref{effective} \cite{Caves1981}.
Squeezed vacuum states have recently been discussed in the context of
non-linear Compton scattering~\cite{DiPiazzaQu2026}, and Breit-Wheeler mechanism \cite{GeZh2026}.
In the process considered here there is no real charge at all --
the signal is sourced by the non-linear response of the polarized
vacuum itself.

As a possible measurement scheme, one could prepare the squeezed 
state~\eqref{single-mode}, let it interact via $\hat U_{\rm eff}$ 
and then reverse the squeezing operation, i.e.,  
$\hat S_\zeta^\dagger\hat U_{\rm eff}\hat S_\zeta\ket{0}$. 
After this sequence, measuring a non-zero photon number amounts to a
detection of the interaction $\hat U_{\rm eff}$. 

To simplify the analysis, we first rotate the phase of $\hat a$ such that
$\zeta>0$ is real.
Note that one should apply the same rotation in Eq.~\eqref{effective}
such that the phase of $\mathfrak A$ changes as well.
Using $\hat S_\zeta^\dagger\hat a\hat S_\zeta=\hat a\cosh\zeta
-\hat a^\dagger\sinh\zeta$ \cite{MandelWolf}, we find
\bea
\hat S_\zeta^\dagger\hat U_{\rm eff}\hat S_\zeta=
\exp\left\{
{\mathfrak A}\left(\hat a ^\dagger\cosh\zeta-\hat a\sinh\zeta
\right)
-\rm h.c.
\right\}.
\label{eq:echo}
\ea
Thus, if ${\mathfrak A}$ is imaginary 
(i.e., if $\zeta$ and ${\mathfrak A}$ are out of phase), 
the amplitude is effectively suppressed
${\mathfrak A}\to{\mathfrak A}\,e^{-\zeta}$.
On the other hand, if ${\mathfrak A}$ is real
(i.e., if $\zeta$ and ${\mathfrak A}$ are in phase),
the amplitude is effectively amplified via
${\mathfrak A}\to{\mathfrak A}\,e^{+\zeta}$.
%
%
%
Assuming random phases between $\zeta$ and ${\mathfrak A}$,
the net effect averages out for small squeezing $|\zeta|\ll1$.
For large squeezing parameters $|\zeta|\gg1$, however, the
amplification dominates on average, see also
Eq.~\eqref{SqRate} below.




In the language of the vacuum-emission picture, the displacement
${\mathfrak A}$ of the single-mode analysis translates to
the vacuum-emission amplitude via ${\mathfrak A}=-iS_{\rm vac}(\f k)$.
Repeating the calculation of Ref.~\cite{KarbsteinShaisultanov2015}
for a mode prepared in a squeezed
vacuum gives
\bea
S_{\rm sq}(\f k)
=
S_{\rm vac}(\f k)
\left[
\cosh|\zeta| - e^{i(\theta-2\phi)}\sinh|\zeta|
\right],
\label{SqAmplitude}
\ea
where $S_{\rm vac}(\f k)=|S_{\rm vac}(\f k)|e^{i\phi}$
and $\zeta=|\zeta|e^{i\theta}$, i.e., considering general phases.
Hence, the ratio of the emitted photon numbers
with and without squeezing reads
%
\bea
\frac{|S_{\rm sq}(\f k)|^2}{|S_{\rm vac}(\f k)|^2}
=1+2n_\zeta
-\,2\sqrt{n_\zeta(n_\zeta+1)}\,\cos(\theta-2\phi),
\qquad
\label{SqRate}
\ea
with $n_\zeta=\sinh^2|\zeta|$ denoting the mean number of photons
in the initial squeezed state.
Due to the interference between the two terms in
Eq.~\eqref{SqAmplitude}, the right-hand side of Eq.~\eqref{SqRate}
varies between $e^{-2|\zeta|}$ and $e^{+2|\zeta|}$, with the maximum
at $\theta-2\phi=\pi$.
The squeezing phase must therefore be matched to \emph{twice} the phase of
the emission amplitude, a consequence of the two-photon nature of
squeezing.
For random phases, the cosine term averages out such that one would
still get an enhancement (as mentioned above).

%

Note that the second squeezing operation applied after the
interaction $\hat U_{\rm eff}$ is essential for this enhancement.
Directly measuring the photon number $\hat n$ after the interaction
$\hat U_{\rm eff}$, i.e., without the second squeezing stage
$\ket{{\rm out}}=\hat U_{\rm eff}\hat S_\zeta\ket{0}$ would
yield the expectation value
\bea
\bra{{\rm out}}\hat n\ket{{\rm out}}
=
n_\zeta+n_{\mathfrak A}
=
\sinh^2|\zeta|+|{\mathfrak A}|^2,
\ea
such that the squeezing would not be beneficial -- but rather a
drawback since it adds a background $\sinh^2|\zeta|$ to the desired
signal $|{\mathfrak A}|^2$.

As another important point, imperfections such as decoherence between
the two squeezing stages could deteriorate the interference mentioned above
and thereby attenuate the signal to noise ratio.
For example, a photon loss $\epsilon$ would introduce an uncanceled background noise of order $\epsilon\sinh^{2}|\zeta|$, which limits
the useful squeezing to $e^{2|\zeta|}<\ord(1/\epsilon)$.
%
%
Let us finally remark that the scheme discussed here is closely related
to $SU(1,1)$ interferometry~\cite{Yurke1986}, see also~\cite{Walls1983}.

\paragraph{Two-mode squeezed state}

As explained above, single-mode squeezing~\eqref{single-mode} corresponds 
to a phase sensitive amplifier, which amplifies the signal 
${\mathfrak A}\to{\mathfrak A}\,e^{+|\zeta|}$ in one direction but
attenuates it ${\mathfrak A}\to{\mathfrak A}\,e^{-|\zeta|}$ in the
orthogonal direction. 
To avoid this strong phase dependence, one could employ a two-mode squeezed 
state \cite{CavesSchumaker1985,SchumakerCaves1985,Yurke1986}
\bea
\label{two-mode}
\ket{\xi}=
\hat S_\xi\ket{0,0}
=
\exp\left\{\xi^*\hat a\hat b-\rm h.c.\right\}
\ket{0,0}
\,,
\ea
which then acts as a phase insensitive amplifier.
%
Here, the phase of $\xi$ can be absorbed by a re-definition of the
idler mode $\hat b$, while the remaining freedom in the phase of
$\hat a$ renders ${\mathfrak A}$ real.


With the same sequence
$\hat S_\xi^\dagger\hat U_{\rm eff}\hat S_\xi\ket{0}$ 
as above, the creation and annihilation operators 
$\hat a^\dagger$ and $\hat a$ contained in $\hat U_{\rm eff}$ 
are modified according to the Bogoliubov transformation 
$\hat S_\xi^\dagger\hat a\hat S_\xi=\hat a\cosh\xi-\hat b^\dagger\sinh\xi$. 
%
%
As a result, the transformed interaction 
$\hat S_\xi^\dagger\hat U_{\rm eff}\hat S_\xi$
corresponds to a coherent displacement in both modes, 
the signal mode $\hat a^\dagger$ and $\hat a$ as well as the idler 
mode $\hat b^\dagger$ and $\hat b$
\bea
\hat S_\xi^\dagger\hat U_{\rm eff}\hat S_\xi=
\exp\left\{
{\mathfrak A}\left(
\hat a^\dagger\cosh\xi-\hat b\sinh\xi
\right)
-\rm h.c.
\right\}. \quad
\ea
As a result, the displacement in the signal mode is enhanced via
${\mathfrak A}\to{\mathfrak A}\,\cosh\xi$ independently of the 
relative phase mentioned above. 
In addition, we also obtain a displacement ${\mathfrak A}\sinh\xi$
in the idler mode which further helps us to detect the vacuum
non-linearity.
Measuring the total photon number in both modes (after reversing
the squeezing operation
$\ket{{\rm out}}=\hat S_\xi^\dagger\hat U_{\rm eff}\hat S_\xi\ket{0}$)
yields
\bea
\bra{{\rm out}}\hat a^\dagger\hat a+\hat b^\dagger\hat b\ket{{\rm out}}
=
|{\mathfrak A}|^2\cosh(2\xi)
\,,
\ea
where the factor $\cosh(2\xi)$ describes the effective amplification
of the amplitude $\mathfrak A$.
As in the case of single-mode squeezing, imperfections between
the two squeezing stages can deteriorate this amplification, where
the case of two-mode squeezing is more robust against phase variations
of $\mathfrak A$ (due to its phase insensitivity).

\paragraph{PVLAS set-up}

Now let us apply the general considerations above to concrete experimental
schemes.
As our first example, we discuss the set-up used by the PVLAS
collaboration.
Translated to our formalism, the classical background field
$A^\mu_{\rm ext}$ consists of a quasi-static magnetic field in the Tesla
regime and a laser field of fixed polarisation with a power of a few mW.
The quasi-static magnetic field contributes quadratically to
$J_\mu^{\rm vac}$
in Eq.~\eqref{effective} while the laser field enters linearly.
As a result, the signal mode has the same wave-number $\f{k}$ as
the incident laser light, but the opposite polarisation.

Comparing the magnetic field in the Tesla regime with the Schwinger
critical (magnetic) field $B_{\rm crit}\approx4\times10^9~\rm T$,
one finds that the vacuum non-linearity in Eq.~\eqref{Euler-Heisenberg}
can be understood as a polarisation dependent change of the refractive
index of order $10^{-23}$.
Placing this small number into context, the required amplification by
23 orders of magnitude is comparable to the detection of
gravitational waves.
Since the number of photons in the incident optical laser is not
enough to achieve such a strong
amplification,
the PVLAS collaboration
used further
enhancement
factors such as a long optical path length
of $\ord(10^6~\rm m)$ and a long integration time (i.e., many repetitions).

%

In this way, the PVLAS collaboration was able to place an upper bound
of order $10^{-22}$ on the polarisation dependent change of the
refractive index at $B_{\rm ext}=2.5$~T whose uncertainty lies one
order of magnitude above the QED prediction~\cite{Ejlli2020}.
Note that PVLAS was ultimately limited by the low-frequency
birefringence noise of the cavity mirrors rather than by shot noise.
Once this technical issue
is controlled, as envisaged in the
successor experiments~\cite{VMBCERN}, the remaining amplification
by one order of magnitude
(in the small-signal limit applicable here)
is well within reach of currently available squeezing \cite{LIGO2013}.

%

\paragraph{All-optical four-wave mixing}

As our next example, let us consider a scheme where the distinction
between the signal mode and the background field $A^\mu_{\rm ext}$
is not just based on polarization, but also on its wave-number $\f{k}$.
One way of realizing this is to combine three ultra-strong and tightly
focused laser pulses with possibly different frequencies which are
incident from three different directions such that their foci overlap
spatially and temporally
\cite{Bernard2000,Lundstrom2006, Gies:2017ezf}.
Merging three real on-shell photons into one photon via the
Euler-Heisenberg Lagrangian~\eqref{Euler-Heisenberg}
is not allowed due to energy-momentum conservation,
but it is possible to absorb two photons from
the background field $A^\mu_{\rm ext}$ and to emit two other photons
-- one of them into the highly occupied background field $A^\mu_{\rm ext}$
and the other one into the signal mode $\hat a^\dagger$ and $\hat a$.

Assuming optical scales in eV regime and ignoring geometrical factors
\cite{King2018}, we obtain the rough scaling law
\bea
\label{four-wave}
{\mathfrak A}\sim\frac{q}{360\pi^2}
\left(\frac{E_L}{E_{\rm crit}}\right)^3
\left(\frac{m}{\omega_L}\right)^2
\ea
in terms of the characteristic field strength $E_L$ and the frequency
$\omega_L=\ord(\rm eV)$ of the strong incident lasers.
Assuming an intensity of order $10^{21}~\rm W/cm^2$ which is reachable
in several laboratories around the world \cite{Danson2019,Yoon2021}
and comparing it with the
Schwinger critical field $E_{\rm crit}$ we find that the small ratio
of field strengths $E_L/E_{\rm crit}\sim10^{-4}$ is roughly compensated
by the large ratio $m/\omega_L\sim10^6$ of the involved energy scales.
However, the pre-factor $q/(360\pi^2)$ from the Euler-Heisenberg
Lagrangian~\eqref{Euler-Heisenberg} still yields a suppression
of four orders of magnitude.
Unless this suppression is compensated by geometrical factors,
we are deep in the small-signal limit.

With currently available squeezing, it is probably very challenging
to compensate the suppression by four orders of magnitude.
However, even a partial amplification via squeezing could
significantly reduce the required repetitions of the experiment.

Increasing the incident intensity to $10^{23}~\rm W/cm^2$ or more,
such that $E_L/E_{\rm crit}\sim10^{-3}$,
we approach the regime where ${\mathfrak A}=\ord(1)$.
Even though this might facilitate a detection of the QED vacuum
non-linearity under ideal conditions, it is highly probable that
one has to face the problem of the stray photons mentioned above.
To resolve this obstacle, one could use coherent states as
initial states, as explained after Eq.~\eqref{addition}.

\paragraph{Coulomb-assisted photon merging}

Since colliding three different laser pulses such that their foci
overlap spatially and temporally is quite challenging, let us now
consider a case with only one optical laser.
To facilitate a non-linear interaction via Eq.~\eqref{Euler-Heisenberg},
we envisage the collision of this incident laser pulse with the Coulomb
field of a nucleus, see also
\cite{Ahmadiniaz2021, Ahmadiniaz:2022mcy, Ahmadiniaz:2023udn}.
In terms of energy-momentum conservation, the Coulomb field can take up
the recoil such that two photons can be absorbed from the incident
laser pulse (photon merging).
Thus the signal photon has twice the frequency of the incident laser
pulse and is emitted in a different direction in general, which makes
it easier to distinguish it from the background.

In this case, 
the Feynman box diagrams underlying the Euler-Heisenberg
Lagrangian~\eqref{Euler-Heisenberg} have three external photon lines
in the optical regime: the two photons absorbed from the incident
laser pulse as well as the emitted signal photon.
Due to energy-momentum conservation, the fourth external photon line
representing the Coulomb field must also have a momentum in the optical
regime (and zero frequency) such that the
low-energy description~\eqref{Euler-Heisenberg} is applicable.
%
This would change for the case of two external Coulomb lines.
Then, 
the momenta need no longer be small and the full off-shell four-photon amplitude is required, see \cite{Ahmadiniaz2023b}.
This applies 
to the well-known cases of Delbr\"uck scattering
\cite{Meitner:1933kww, Jarlskog:1973aui} or the ATLAS experiment
\cite{dEnterria:2013zqi, ATLAS:2017fur} where two real photons
were created by the interaction of two ``virtual'' photons representing
the Coulomb fields of nuclei in ultra-peripheral collisions.
Hence, even though both phenomena can also be interpreted in terms of
the vacuum non-linearity, they refer to ultra-short distance  scales
instead of the macroscopic (long-distance) non-linearity considered here.


Using the low-energy approach, 
we may perform an analogous estimate as in
Eq.~\eqref{four-wave} and arrive at
\bea
\label{Coulomb}
{\mathfrak A}\sim\frac{Zq^3}{360\pi^2}
\left(\frac{E_L}{E_{\rm crit}}\right)^2,
\ea
where $Z$ denotes the atomic number of the nucleus.
Assuming the same optical intensity of order $10^{21}~\rm W/cm^2$
as before, we see that the small ratio of field strengths
$E_L/E_{\rm crit}\sim10^{-4}$ is not compensated by the large
energy ratio $m/\omega_L\sim10^6$ anymore. 
Thus, obtaining a detectable signal from a single nucleus is
probably very hard.
On the other hand, if we consider a large number $N$ of
(fully ionized) nuclei, e.g., $N\sim10^8$ \cite{Ahmadiniaz2021},
which are
close enough together such that their amplitudes~\eqref{Coulomb}
add up coherently, this small ratio of field strengths
$E_L/E_{\rm crit}\sim10^{-4}$ could be compensated.

Nevertheless, we are still in the small-signal limit due to the
pre-factor.
This again motivates amplifying the amplitude~\eqref{Coulomb}
by sending in squeezed states.
Note that one could also consider effectively enhancing the
relevant field strength $E_L$ in the rest frame of the nuclei
by colliding the optical laser with a counter-propagating
beam of relativistic nuclei,
see also \cite{DiPiazza2008b,GieKarbSha2014}.
However, ensuring the coherent superposition of the amplitudes
from a sufficiently large number of nuclei would then be more
challenging.

\paragraph{Conclusions}

We study how the detection of the QED vacuum non-linearity could be
facilitated by sending in excited photon states instead of the vacuum
state.
A coherent (i.e., classical) state can help to distinguish the signal
from noise such as stray photons while a squeezed (i.e., non-classical) state
can even amplify the signal -- provided that squeezing stages before
{\em and} after the interaction are applied.
For three example scenarios, the PVLAS set-up, all-optical four-wave
mixing and coulomb-assisted photon merging, we discuss the application
of these general considerations.

As an outlook, one could also consider vacuum birefringence with
x-rays, see, e.g., \cite{Heinzl2006, AhmadiniazDetection2023, DiPiazza:2006pr}.
Although the general ideas apply in basically the same way,
interferometry and, even more so, squeezing with x-rays is
typically more challenging than in the optical regime.
Note, however, that the dark-field scheme currently pursued in
an experimental campaign at HIBEF \cite{BIREF-HIBEF2025, PhysRevD.109.096009, Smid2025}
can be interpreted as an interference effect.
Apart from the QED vacuum non-linearity, axions would generate
very analogous effects
${\cal L}_{\rm int}^{\rm eff}\propto(\f{E}\cdot\f{B})^2$
and thus the same ideas can be applied in this context \cite{Dobrich:2010hi, Villalba-Chavez:2013bda,Evans:2023jpr}.

\appendix

\section{Appendix: Non-Gaussian states}

\paragraph{Fock state}

As another example for a non-classical photonic state, let us consider
the Fock state $\ket{{\rm in}}=\ket{n}$.
This state with a well-defined initial photon number does also allow us
to effectively amplify the signal.
In the small-signal limit, the final state reads
\bea
\label{Fock}
\ket{{\rm out}}=
\ket{n}+\sqrt{n+1}\,{\mathfrak A}\ket{n+1}-
\sqrt{n}\,{\mathfrak A}^*\ket{n-1},
\ea
up to small corrections $\ord({\mathfrak A}^2)$.
As a result, the amplitude is effectively enhanced by
$\ord(\sqrt{n})$ for large $n$.
In principle $\ket{n}$ can be distinguished with certainty from
$\ket{n\pm1}$, but doing that in practice (especially for large $n$),
e.g., by photonic state tomography \cite{LvovskyRaymer2009},
is non-trivial.
In addition, preparing an initial Fock state with large $n$
is typically harder than generating a coherent state
(which is the natural output of a laser, for example) \cite{Cooper2013}.
Note that the enhancement $\ord(n)$ in the probability is comparable
to that obtained from a squeezed state containing the same mean photon
number $n_\zeta=\sinh^2|\zeta|$, the practical difference lies in the
preparation and the read-out.


\paragraph{NOON state}

In analogy to the step from single-mode squeezing to two-mode squeezing,
we may also consider the NOON state~\cite{Boto2000,Dowling2008}
\bea
\label{noon}
\ket{{\rm in}}=\frac{\ket{n,0}+\ket{0,n}}{\sqrt{2}}=
\frac{(\hat a^\dagger)^n+(\hat b^\dagger)^n}{\sqrt{2\,n!}}\,\ket{0,0},
\ea
where we again incorporated an idler mode $\hat b$.
As for the Fock state, the interaction amplitude is effectively enhanced
by $\ord(\sqrt{n})$ for large $n$ (in the small-signal limit).
%
%
However, the generation of photonic NOON states with large $n$
as well as NOON state interferometry are very challenging.



\paragraph{Schr\"odinger cat state}

As our final example, let us consider a Schr\"odinger cat state
of the form
\bea
\label{cat}
\ket{{\rm in}}=\frac{\ket{\alpha,0}+\ket{0,\alpha}}{\sqrt{2}}=
\frac{\hat D_{\hat a}(\alpha)+\hat D_{\hat b}(\alpha)}{\sqrt{2}}
\,\ket{0,0},
\ea
where $\hat D_{\hat a}(\alpha)$ and $\hat D_{\hat b}(\alpha)$
are coherent displacement operators for the signal $\hat a$
and the idler $\hat b$ mode, respectively.
For large $\alpha$, the two states $\ket{\alpha,0}$
and $\ket{0,\alpha}$ are almost orthogonal, i.e.,
macroscopically distinct (in analogy to the cat being
either dead or alive).
Now, if we operate in the case where the phases of $\alpha$
and $\mathfrak A$ are shifted by $\pi/2$ (e.g., one is real
and one imaginary), such that the interaction~\eqref{effective}
basically corresponds to a phase shift, we see that an amplitude
$\mathfrak A=\ord(1/\alpha)$ is sufficient to transform the
initial state~\eqref{cat} to the final state
\bea
\label{cat-out}
\ket{{\rm out}}=\frac{\ket{\alpha,0}-\ket{0,\alpha}}{\sqrt{2}}
\,.
\ea
Again for large $\alpha\gg1$, this state is nearly orthogonal
to the initial state~\eqref{cat} and thus can in principle be
distinguished with almost certainty by suitable tomographic
measurements -- although doing this in practise is highly non-trivial.
Very analogous to the difference between the Poisson limit and the
Heisenberg limit in quantum sensing, the required amplitude
$\mathfrak A$ is reduced by a factor of $\ord(\alpha)$, i.e.,
$\ord(\sqrt{n})$.
In summary, the three examples of non-Gaussian states~\eqref{Fock},
\eqref{noon} and \eqref{cat} considered above do basically show that
same $\ord(\sqrt{n})$ enhancement of the amplitude $\mathfrak A$
as the squeezed states~\eqref{single-mode} and \eqref{two-mode}.

\end{document}